\documentstyle[aps,psfig,twocolumn]{revtex}
\begin{document}
\title{Magnons in real materials from density-functional theory}
\author{Ralph Gebauer$^{1}$\cite{ralph} and Stefano
Baroni$^{1,2}$\cite{stefano}}

\address{$^1$CECAM -- Centre Europ\'een de Calcul Atomique et
Mol\'eculaire, ENSL, 46 All\'ee d'Italie, F-69364 Lyon Cedex 07,
France \\ $^2$INFM -- Istituto Nazionale di Fisica della Materia and
\\ SISSA -- Scuola Internazionale Superiore di Studi Avanzati, via
Beirut 2/4, I-34014 Trieste, Italy } \date{\today}

\maketitle

\begin{abstract} We present an implementation of the {\it adiabatic
spin-wave dynamics} of Niu and Kleinman. This techniques allows to
decouple the spin and charge excitations of a many-electron system
using a generalization of the adiabatic approximation. The only input
for the spin-wave equations of motion are the energies and Berry
curvatures of many-electron states describing frozen spin spirals. The
latter are computed using a newly developed technique based on
constrained density-functional theory, within the local spin density
approximation and the pseudo-potential plane-wave method. Calculations
for iron show an excellent agreement with experiments. \end{abstract}

\pacs{75.30.Fv, 
      71.15.Mb, 
      75.40.Gb, 
      75.50.Bb  
     }

\narrowtext

Over the last 30 years, Density-Functional Theory (DFT) \cite{dft} has
brought a large number of properties of real materials within the
predictive range of first-principles calculations. Even though excited
states cannot be properly described within DFT---which strictly
speaking only applies to the electronic ground states---the use of the
adiabatic approximation allows to predict the vibrational excitation
spectra of semiconductors \cite{pasqua} and metals \cite{sdg} to a
very high accuracy. Magnetic excitations are also adiabatically
decoupled from charge excitations. However, the adiabatic decoupling
of spin excitations has only been implemented so far using {\it
ad-hoc} models---such as the Heisenberg Hamiltonian
\cite{russi}---which are difficult to reconcile with an itinerant
picture of magnetism. A substantial step toward the calculation of
magnetic excitations within a firm theoretical framework has been done
by Niu and Kleinman (NK) \cite{niu} who showed how a generalization of
the Born-Oppenheimer (BO) method can be used to rigorously, although
approximately, decouple spin and charge excitations of a many-electron
system. The NK method lends itself quite naturally to be implemented
within DFT. The latter requires a proper description of non-collinear
magnetic structures, a task which has hardly been tackled so far in
full generality. Although the local spin-density approximation (LSDA)
was formulated long ago so as to account for general magnetic
structures \cite{vonbarth}, the vast majority of the LSDA calculations
assumes that the magnetization is aligned to a same direction all over
the system. Most of the existing applications to non collinear
structures rely on some sort of atomic-sphere approximation (ASA) in
which different spin quantization axes are chosen within different
spheres \cite{kuebler}. Progresses beyond this rather crude
approximation have been made only recently \cite{singh,car,o2}.

In the present work we combine the adiabatic decoupling method of
Ref. \cite{niu} with a newly developed {\it constrained DFT scheme}
which allows the calculation of the spin-spiral states necessary to
implement the NK spin-wave equations of motion, without using any
atomic-sphere approximation. Our implementation---which is based on
the plane-wave pseudo-potential method---is successfully demonstrated
by calculating the magnon dispersions of iron. This paper is organized
as follows: we first briefly review and generalize the NK adiabatic
decoupling scheme; we then introduce our constrained DFT approach to
spin-spiral states, and we finally present our results for iron.

In order to establish an adiabatic decoupling scheme that allows to
derive from first principles the spin-fluctuation equations of motion,
we first derive the classical equations of motion for nuclei in a
molecule or in a solid by using of the concept of {\it adiabatic
coherent state}. This procedure has the advantage of not making
explicit reference to the {\it smallness} of the nuclear kinetic
energy term in the Hamiltonian. Because of this, we can then make the
ansatz that a similar method applies in the general case where no such
small term exists in the Hamiltonian, but the density of states of the
system is still characterized by a low-lying portion which is well
separated in energy from the rest. This procedure provides thus an
unified approach to the classical BO approximation and to the NK
method.

Let us consider a system of interacting electrons (of mass $m=1$) and
nuclei (of mass $M \gg m$ and in number $N$), and let us indicate by
capital letters the coordinates and momenta of the nuclei,
($\bbox{R},\bbox{P}$), and by lower-case letters those of the
electrons, ($\bbox{r},\bbox{p}$). The Hamiltonian of the system reads:
\begin{equation} \bbox{H} = {1\over 2 M} \bbox{P}^2 + {1\over 2}
\bbox{p}^2 + V(\bbox{r},\bbox{R}) \label{hamiltonian} \end{equation}
(here and in the following quantum operators are indicated by
boldfaces). Let us now consider the wave-function $\Psi_{\cal
R,P}(r,R)$ which minimizes the expectation value of the Hamiltonian,
Eq.  (\ref{hamiltonian}), with the constraint that the expectation
values of $\bbox{R}$ and $\bbox{P}$ equal $\cal R$ and $\cal P$,
respectively. By introducing the $\lambda$ and $\mu$ Lagrange
multipliers for the constraints on $\bbox{R}$ and $\bbox{P}$
respectively, $\Psi_{\cal R,P}$ can be formally obtained as the
ground-state solution of the eigenvalue equation: \begin{equation}
\bigl [ \bbox{H} - \lambda (\bbox{R} - {\cal R} ) - \mu (\bbox{P} -
{\cal P} ) \bigr ] \Psi_{\cal R,P} = \epsilon({\cal R,P}) \Psi_{\cal
R,P}, \label{constrained_eq} \end{equation} where $\lambda \equiv
\lambda({\cal R,P})$ and $\mu \equiv \mu({\cal R,P})$ are chosen in
such a way that: $ \langle \Psi_{\cal R,P} | \bbox{R} | \Psi_{\cal
R,P} \rangle = {\cal R} $ and $ \langle \Psi_{\cal R,P} | \bbox{P} |
\Psi_{\cal R,P} \rangle = {\cal P} $. It is easy to verify that when
$\bbox{H}$ is the Hamiltonian of a harmonic oscillator, $\bbox{H}=
{1\over 2 M} \bbox{P}^2 + {1\over 2} k \bbox{R}^2$, the eigenvalue
equation, Eq. (\ref{constrained_eq}), yields directly Glauber's
canonical coherent states \cite{glauber}. For this reason, in the
general case we name $ \Psi_{\cal R,P}(r,R) $ an {\it adiabatic
coherent state} (ACS) \cite{coh-sta} and we refer to $({\cal R},{\cal
P})$ as to the {\it label} of the ACS.

Let us suppose that the time evolution of an ACS is an ACS: this is
true for harmonic oscillators and canonical coherent states, but in
general this is only an approximation. The low energy dynamical
properties of the system are then given by the variational problem:
\begin{equation} \delta \int_{t_0}^{t_1} \left \langle \Psi_{{\cal
R}(t),{\cal P}(t)} \left | i{\partial \over \partial t} - \bbox{H}
\right | \Psi_{{\cal R}(t),{\cal P}(t)} \right \rangle dt =0,
\label{variational_principle} \end{equation} which, following the NK
procedure, leads the equations of motion for the $({\cal R,P})$ label:
\begin{equation} \left( \begin{array}{rr}  \Omega^{\cal R,R}  &
\Omega^{\cal R,P} \\ \Omega^{\cal P,R} & \Omega^{\cal P,P}
\end{array} \right) \left( \begin{array}{c} \dot{{\cal R}} \\
\dot{{\cal P}} \end{array} \right) = \left( \begin{array}{c}
\frac{\partial \epsilon({\cal R,P})}{\partial {\cal R}} \\
\frac{\partial \epsilon({\cal R,P})}{\partial {\cal P}} \end{array}
\right), \label{matrix} \end{equation} where the $\Omega$'s are
matrices defined as $ \Omega^{\cal A,B}_{i,j} = -2\ {\rm Im} \left
\langle {\partial \Psi_{\cal R,P} \over \partial {\cal A}_i} \left |
{\partial \Psi_{\cal R,P} \over \partial {\cal B}_j} \right \rangle
\right .$, and the $i$ or $j$ subscripts indicate the components of
the $\cal R$ and $\cal P$ $3N$-dimensional vectors. From this
definition, it follows that $ \Omega^{\cal R,P}_{i,j} = - \Omega^{\cal
P,R}_{j,i} $. Following the literature \cite{niu,MBO}, we name the
$\Omega$ matrices the {\it Berry curvature} of the ACS.  In order to
recover the classical equations of motion for the nuclear degrees of
freedom, we first simplify the eigenvalue equation
(\ref{constrained_eq}) using the fact that $\bbox{H}$ is quadratic in
$\bbox{P}$. Taking into account that $[\bbox{P},{\rm e}^{i{\cal
P}\bbox{R}}] = {\cal P}\ {\rm e}^{i{\cal P}\bbox{R}}$, a simple
algebraic manipulation in Eq.~(\ref{constrained_eq}) allows to cast
$\Psi_{\cal R,P}$ and $\epsilon({\cal R,P})$ into the form:
\begin{eqnarray} \Psi_{\cal R,P}(r,R) & = & {\rm e}^{i{\cal
P}R}\Phi_{\cal R}(r,R), \label{psi-phi} \\ \epsilon({\cal R,P}) & = &
{1\over 2M} {\cal P}^2 + \epsilon({\cal R}), \label{epsilon}
\end{eqnarray} where the real wave-function  $\Phi_{\cal R}$ satisfies
the eigenvalue equation: \begin{equation} \bigl [ \bbox{H} - \lambda
(\bbox{R} - {\cal R} ) \bigr ] \Phi_{\cal R} = \epsilon({\cal R})
\Phi_{\cal R}, \label{real_constrained_eq} \end{equation} and the
Lagrange multiplier $\lambda \equiv \lambda({\cal R})$ is chosen in
such a way that: $ \langle \Phi_{\cal R} | \bbox{R} | \Phi_{\cal R}
\rangle = {\cal R} $. Using Eq. (\ref{psi-phi}) and the reality of
$\Phi_{\cal R}$, the $\Omega$'s can be explicitly calculated:
$\Omega^{\cal R,P}_{i,j} = -2\ \left \langle \left . \left . {\partial
\Phi_{\cal R} \over \partial {\cal R}_i} \right | \bbox{R}_j \ \right
| \Phi_{\cal R} \right \rangle = -{\partial \over \partial {\cal R}_i}
\langle \Phi_{\cal R} | \bbox{R}_j | \Phi_{\cal R} \rangle =
-\delta_{i,j}$, and $\Omega^{\cal R,R}_{i,j} = \Omega^{\cal P,P}_{i,j}
= 0$. By inserting these $\Omega$'s and the expression given by
Eq. (\ref{epsilon}) for $\epsilon({\cal R,P})$ into
Eq. (\ref{matrix}), one finally arrives at a set of Hamilton-like
equations for the nuclear motion. In order identify $\epsilon({\cal
R})$ with the familiar BO energy surface, it is enough to neglect in
Eq. (\ref{real_constrained_eq}) the nuclear kinetic-energy term
present in $\bbox{H}$ ($1/M \approx 0$). It is important to remark
that---while the smallness of $1/M$ is the physical reason why nuclear
vibrations are adiabatically decoupled from electronic
excitations---this property has not been formally used to establish
the classical equations of motion, but only to identify
$\epsilon({\cal R})$ with the traditional BO energy surface. In
principle, $\epsilon({\cal R})$ could be as well calculated from its
definition, Eq. (\ref{real_constrained_eq}), without letting $1/M$ go
to $0$. When studying the electronic spectrum of magnetic systems, no
such small term exists in the Hamiltonian. Nevertheless, spin
excitations can be adiabatically decoupled from charge excitations by
using the {\it magnetic BO surface} obtained from the magnetic analog
of Eq. (\ref{real_constrained_eq}).

Spin excitations are characterized by the fluctuation of the spin
polarization, $\vec {\cal M}(\vec r)$,
much in the same way as nuclear vibrations are
characterized by the fluctuation of the nuclear canonical coordinates,
$({\cal R,P})$. In both cases, the fluctuating observable can be
identified with the label of an appropriate ACS. Magnetic ACS (MACS)
are thus labeled by a vector field $\vec {\cal M}(\vec r)$, $
\Psi_{[\vec {\cal M}]}, $ and they are solution of the variational
problem: \refstepcounter{equation} \label{varmag-1}
$$\displaylines{\hfill E[\vec {\cal M}] = {\rm min} \langle \Psi |
\bbox{H} | \Psi \rangle , \hfill (\theequation) \cr \hbox{\rm with the
constraint:} \hfill \cr \refstepcounter{equation} \label{varmag-2}
\hfill \langle \Psi | \bbox{\psi}(\vec r)^+ \vec \sigma
\bbox{\psi}(\vec r) | \Psi \rangle = \vec {\cal M}(\vec r), \hfill
(\theequation) } $$ where $\bbox{H}$ is the many-body Hamiltonian of
the system, the $\psi$'s are fermion spinor field operators, and
$\vec\sigma \equiv (\sigma_x,\sigma_y,\sigma_z)$ the Pauli matrices.

Following NK, the linearized equations of motion for $\vec {\cal
M}(\vec r,t)$ read ($\hbar=1$): \refstepcounter{equation}
\label{lin-eq-mot} $$ \displaylines{\quad \sum_\beta \int d^3 r\ ''
\Omega_{\alpha \beta}( r\ ',\vec r\ '') \dot {\cal M}_\beta(\vec r\
'',t) = \hfill \cr \hfill \sum_\beta \int d^3 r\ '' K_{\alpha
\beta}(\vec r\ ',\vec r\ '') {\cal M}_\beta(\vec r\ '',t), \quad
(\theequation) } $$ where $\Omega$ is the Berry curvature of the MACS
and $K$ the matrix of the second derivatives of its energy with
respect to its label, $\vec {\cal M}$: \begin{eqnarray}
&\Omega_{\alpha\beta}(\vec r\ ',\vec r\ '') = - 2\ {\rm Im} \left
\langle {\delta \Psi[\vec {\cal M}] \over \delta {\cal M}_\alpha(\vec
r\ ') } \right | \left .  {\delta \Psi[\vec {\cal M}] \over \delta
{\cal M}_\beta(\vec r\ '')} \right \rangle, & \label{omega} \\ &
K_{\alpha\beta}(\vec r\ ',\vec r\ '') = {\delta^2 E[{\cal M}] \over
\delta{\cal M}_\alpha(\vec r\ ') \delta{\cal M}_\beta(\vec r\
'')}. \label{kappa} \end{eqnarray} In order to calculate the Berry
curvature, we use a finite-difference approach \cite{niu,rapix}. To
keep the notation as simple as possible, we define $\Psi_0 \equiv
\Psi[{\cal M}_\alpha(\vec r\ ')]$, $\Psi_1 \equiv \Psi[{\cal
M}_\alpha(\vec r) + \epsilon \delta_{\alpha\beta} \delta(\vec r-\vec
r\ ')]$, and $\Psi_2 \equiv \Psi[{\cal M}_\alpha(\vec r) + \epsilon
\delta_{\alpha\gamma} \delta(\vec r-\vec r\ '')]$. We have then:
$\Omega_{\alpha\beta}(\vec r\ ',\vec r\ '') \approx -{2\over
\epsilon^2}\ {\rm Im} \langle \Psi_1-\Psi_0 | \Psi_2-\Psi_0 \rangle =
-{2\over \epsilon^2}\ {\rm Im} ( \langle \Psi_1| \Psi_2 \rangle +
\langle \Psi_2| \Psi_0 \rangle + \langle \Psi_0| \Psi_1 \rangle)$. By
using the relation: $\langle \Psi_1|\Psi_2 \rangle \approx {\rm log}
\langle \Psi_1|\Psi_2 \rangle +1$, we finally arrive at the relation:
$ \Omega_{\alpha
\beta}(\vec r\ ',\vec r\ '') = -{1\over \epsilon^2}
{\rm Im~log}\ ( \langle \Psi_0| \Psi_1 \rangle \langle \Psi_1| \Psi_2
\rangle \langle \Psi_2| \Psi_0 \rangle) $.

Before proceeding further, we transform the constrained variational
problem, Eqs. (\ref{varmag-1},\ref{varmag-2}), into a non-constrained
one by making use of a penalty functional, $P[\Psi]$, which favors
those states whose magnetization, $\vec M_{[\Psi]}(\vec r) \equiv
\langle \Psi | \bbox{\psi}(\vec r)^+ \vec \sigma \bbox{\psi}(\vec r) |
\Psi \rangle $, is very close to the MACS label, $ \vec {\cal M}(\vec
r) $: $ P[\Psi,\vec {\cal M}] = A \int (\vec M_{[\Psi]}(\vec r) - \vec
{\cal M}(\vec r) )^2 d^3r $. In the limit where $A\to +\infty$, the
state which minimizes the functional: $ E[\vec {\cal
M},\Psi] = \langle \Psi |\bbox{H}| \Psi \rangle + A \int (\vec
M_{\Psi}(\vec r) - \vec {\cal M}(\vec r) )^2 d^3r $
without any constraints but the trivial one on state normalization,
will solve the variational problem, Eq.  (\ref{varmag-1}), with the
constraint given by Eq. (\ref{varmag-2}).  Variation with respect to
$\Psi^*$ leads to the eigenvalue equation: \refstepcounter{equation}
\label{pen-schr} $$\displaylines{\quad \bbox{H} \Psi + 2A \sum_i
\vec\sigma_i \cdot ( \vec M_{[\Psi]}(\vec r_i) - \vec {\cal M}(\vec
r_i)) \Psi = \hfill \cr \hfill E_A[{\cal M}] \Psi, \quad
(\theequation) } $$ where the sum runs over electrons.

Eq. (\ref{pen-schr}) is formally equivalent to a many-body
Schr\"odinger equation where, in addition to physical interactions,
electrons are subject to an external magnetic field whose magnitude,
$\vec B(\vec r) = 2A \left (\vec M_{[\Psi]}({\vec r}) - \vec {\cal M}
({\vec r}) \right )$ depends self-consistently upon the solution of
the equation, $\Psi$. When $A$ grows large, $\vec B$ has the effect to
make $\vec M_{[\Psi]}(\vec r)$ closer and closer to ${\cal M}(\vec
r)$, so that $B(\vec r)\equiv 2 A( \vec M_{[\Psi]}({\vec r}) - \vec
{\cal M}({\vec r})$ tends to a finite limit, and the value of the
penalty functional tends to zero. We conclude that in the large-$A$
limit the ground-state energy of the Schr\"odinger equation,
Eq. (\ref{pen-schr}), gives the value of the constrained minimum,
Eqs. (\ref{varmag-1},\ref{varmag-2}): $\lim_{A\to +\infty} E_A[\vec
{\cal M}] = E[\vec {\cal M}]$.

An approximate solution of Eq. (\ref{pen-schr}) can be found using DFT
within the LSDA, provided that the latter is formulated in such a way
as to allow for arbitrary, non collinear, magnetic structures. In the
LSDA, the many-body Hamiltonian appearing in Eq. (\ref{pen-schr}) is
replaced with an independent-electron Hamiltonian where the effective
one-body potential depends self-consistently upon the electron charge
density and spin polarization. The implementation of the constraining
penalty function in a non-collinear LSDA code thus requires a trivial
modification of the dependence of the self-consistent potential upon
the spin polarization.

Inspection of the microscopic distribution of the magnetization
resulting from various LSDA calculations shows that it is
non-vanishing mostly in the neighborhood of atoms (or molecules, in
magnetic molecular crystals), and that within these neighborhoods its
direction remains almost constant even when different atomic
(molecular) moments are not aligned. At variance with the assumptions
which underlie the ASA treatment of (non-collinear) magnetism, small
deviations from perfect alignment may occur even within an atomic (or
molecular) volume, and this quasi-collinearity results from the
contribution of many occupied Kohn-Sham orbitals whose individual
magnetization is far from collinear. In view of this fact, we choose
to label the MACS with atomic (or molecular) moments, rather than with
a continuous magnetization. This discretization introduces a natural
coarse graining in the treatment of the excitation spectrum, by
cutting away all the fluctuations whose wavelength is shorter than the
inter-atomic (-molecular) distance, and whose high-energies would be
incompatible with the adiabatic decoupling scheme adopted here.
Atomic magnetic moments are defined through the relation: $
\vec{m}(\vec R) = \int \mbox{d}\vec{r} \, \, w(\vec{r} - \vec{R})
\,\vec{\cal M} (\vec{r})$, where the $\vec{R}$'s indicate atomic
positions, and $w(\vec{r})$ is a {\it weight function} which is equal
to one in a region around the origin and zero elsewhere. In our
applications, we choose these regions to be the largest possible
non-overlapping spheres. Within this picture, long-wavelength magnetic
excitations can be thought as time-dependent deviations of the
localized moments from their ground-state configuration. Analogously
to the current terminology adopted for vibrational excitations, a
pattern of atomic moments, $\{ \vec{m}(\vec R) \} $ is called a {\it
frozen magnon}, and the corresponding matrix of second derivatives of
the energy, $\partial^2 E / \partial m_\alpha(\vec R) \partial
m_\beta(\vec R')$, the matrix of the {\it inter-atomic torque
constants}. The self-consistent equations which yield the MACS
corresponding to a frozen magnon have a form similar to that of
Eq. (\ref{pen-schr}), where the self-consistent constraining magnetic
field is different from zero only within the atomic volumes. For any
given wave-vector of the magnon, this equation can be solved using
super-cell techniques, much in the same way as these are applied to
phonon calculations.

If one restricts frozen magnons to configurations described by {\it
magnetic spirals} \cite{Herring}, then the corresponding MACS can be
calculated without any super-cell, by using a generalized Bloch
theorem \cite{Herring,Uhl}.  The magnetic configuration of a spiral of
wave-vector $\vec q$ is such that atomic moments of fixed magnitude
form an angle $\theta$ with respect to the $z$ direction, independent
of $\vec R$, while the polar angle of $\vec m(\vec R)$ has the form:
$\phi(\vec R) = \phi(0) + \vec q \cdot \vec R$.

In order to work out the spin-wave equations of motion, let us
consider the special case of a mono-atomic ferro-magnetic crystal,
such as bcc iron. In our applications, we constrain $\theta$, while
the magnitude of the atomic moments is left unconstrained, and it
turns out to be independent of $\theta$ for small angles, thus giving
an a-posteriori justification of the spiral model. Translational
invariance implies that $\Omega$ and $K$ depend on $\vec R$ and $\vec
R'$ only through their differences; rotational invariance around the
$z$ axis implies that ${K}_{xy} = {K}_{yx} = 0$ while the fact that
${\Omega}_{xx} = {\Omega}_{yy} = 0$ follows from their definition,
Eq. (\ref{omega}). Using these relations, the equation of motion for
the localized moments can be cast into the form: $ -i \tilde
{\Omega}_{xy} (\vec{q}) \, \frac{\mbox{d}}{\mbox{d} t} \bar{m}
(\vec{q}) = \tilde{K}_{xx} (\vec{q}) \, \bar{m} (\vec{q}) $, where
$\tilde A(\vec q)$ indicates the Fourier transform of $A(\vec R)$, and
$\bar{m} (\vec{q}) = \tilde m_x (\vec{q}) + i \tilde m_y (\vec{q})$.
Magnon frequencies are thus given by: $\omega(\vec q) = \tilde
K_{xx}(\vec q) / \tilde \Omega_{xy}(\vec q)$.

We have applied this method to the calculation of magnon frequencies
in bcc iron. The computations are performed using the LSDA \cite{PZ}
within the plane-wave pseudo-potential method \cite{tm}. Plane waves
up to a kinetic-energy cutoff of 90 Ry have been included in the basis
set, while Brillouin-zone integrations were performed on a $8\times
8\times 8$ mesh, with a Gaussian-smearing of 0.03 Ry \cite{GS}.  In
Fig. 1 we display the results of our calculations, performed for $\vec
q$ along the (100) direction, while the ground-state magnetization
axis was chosen to be (001). The agreement with available experiments
is very good, thus giving confidence in the general theoretical
framework followed in this work.

We believe that the adiabatic decoupling scheme by Niu and Kleinman,
based on the concept of adiabatic coherent state, still needs stronger
theoretical foundations. However, we have demonstrated that this
scheme naturally leads to the well established Born-Oppenheimer
classical equation of motion for nuclei in molecules or
solids. Furthermore, this scheme can be straightforwardly implemented
within density-functional theory and, when applied to magnetic
excitations of bcc iron, it leads to magnon dispersions in excellent
agreement with experiment. We conclude that further theoretical and
computational work will probably demonstrate the effectiveness of this
method to deal with slow dynamics in quantum systems.

\newpage

\begin{figure} \psfig{figure=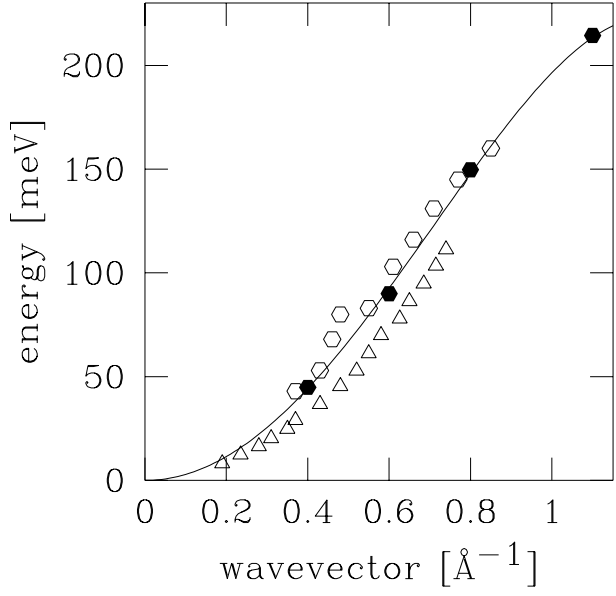,width=8.5cm} 
\caption{Calculated and experimental magnon dispersions in iron. Open
hexagons: pure iron at 10K [20]. Open triangles: Fe
with 12 \% Si) at room temperature [21]. Full
hexagons: this work. The line is a guide for the eye. }

\end{figure}


\begin{references}
\bibitem{rapix} The present derivation of the expression for $\Omega$
is due to R. Resta.
\end{references}
\end{document}